\documentclass[aps,preprint,preprintnumbers,nofootinbib,superscriptaddress,tightenlines,byrevtex,showpacs]{revtex4}
\usepackage{amsmath,amssymb,amsbsy}
\usepackage{graphicx,subfigure}
\usepackage{comment}

\def\str{{\mathrm{str}}}

\begin{document}

\unitlength=1mm

\def\a{{\alpha}}
\def\b{{\beta}}
\def\d{{\delta}}
\def\D{{\Delta}}
\def\e{{\epsilon}}
\def\g{{\gamma}}
\def\G{{\Gamma}}
\def\k{{\kappa}}
\def\l{{\lambda}}
\def\L{{\Lambda}}
\def\m{{\mu}}
\def\n{{\nu}}
\def\w{{\omega}}
\def\O{{\Omega}}
\def\S{{\Sigma}}
\def\s{{\sigma}}
\def\t{{\tau}}
\def\th{{\theta}}
\def\x{{\xi}}

\def\ol#1{{\overline{#1}}}

\def\Dslash{D\hskip-0.65em /}
\def\dslash{{\partial\hskip-0.5em /}}
\def\vslash{{\rlap \slash v}}
\def\qbar{{\overline q}}

\def\CPT{{$\chi$PT}}
\def\QCPT{{Q$\chi$PT}}
\def\PQCPT{{PQ$\chi$PT}}
\def\tr{\text{tr}}
\def\str{\text{str}}
\def\diag{\text{diag}}
\def\order{{\mathcal O}}
\def\vit{{\it v}}
\def\vD{\vit\cdot D}
\def\am{\alpha_M}
\def\bm{\beta_M}
\def\gm{\gamma_M}
\def\smb{\sigma_M}
\def\smt{\overline{\sigma}_M}
\def\tb{{\tilde b}}

\def\mc#1{{\mathcal #1}}

\def\Bbar{\overline{B}}
\def\Tbar{\overline{T}}
\def\cBbar{\overline{\cal B}}
\def\cTbar{\overline{\cal T}}
\def\pq{(PQ)}

\def\eqref#1{{(\ref{#1})}}

%
%
\newcount\hour \newcount\hourminute \newcount\minute 
\hour=\time \divide \hour by 60
\hourminute=\hour \multiply \hourminute by 60
\minute=\time \advance \minute by -\hourminute
\newcommand{\mydate}{\ \today \ - \number\hour :\number\minute}

%
%
\preprint{UMD-40762-393}

\title{\bf Effective Field Theory for the Anisotropic Wilson Lattice Action}

\author{Paulo F. Bedaque}
\email[]{bedaque@umd.edu} 
\affiliation{Maryland Center for Fundamental Physics\\
	Department of Physics, University of Maryland,
	College Park, MD 20742-4111}

\author{Michael I. Buchoff}
\email[]{mbuchoff@umd.edu}
\affiliation{Maryland Center for Fundamental Physics\\
	Department of Physics, University of Maryland,
	College Park, MD 20742-4111}

\author{ Andr\'e Walker-Loud} 
\email[]{walkloud@umd.edu}
\affiliation{Maryland Center for Fundamental Physics\\
	Department of Physics, University of Maryland,
	College Park, MD 20742-4111}

\date{\mydate}
%
%
\begin{abstract}
We construct the effective field theory appropriate for describing the low energy behavior of anisotropic Wilson lattice actions and the $\mc{O}(a)$ improved variant thereof.  We then apply this effective field theory to the hadron spectrum and dispersion relations, focussing on the corrections due to the anisotropy.  We point out an important feature of anisotropic lattices regarding the Aoki-regime; for a given set of fermion masses and spatial lattice spacing, if an isotropic action is in the QCD-phase, this does not guarantee that the anisotropic action is outside the Aoki-regime.  This may be important in the tuning of bare parameters for anisotropic lattices using domain-wall and overlap fermions as well as Wilson and 
$\mc{O}(a)$-improved Wilson fermions.
\end{abstract}

\maketitle

%

%
%
\section{Introduction}

Lattice QCD calculations are necessarily performed with a finite lattice spacing, at a finite spatial extent and presently, with unphysically large quark masses.  Thus, in order to obtain continuum, infinite volume  results at the physical quark masses, multiple extrapolations are required.  At large pion masses, naive (usually polynomial) extrapolations of results obtained with different quark masses and lattice spacings are performed. However, for results computed with quark masses in the chiral regime (corresponding to pion masses on the order of 400 MeV or lower), a generalization of chiral perturbation theory, which includes lattice spacing effects, can provide the analytic form of the dependence of many observables with the lattice spacing and quark mass \cite{Creutz:1996bg,Sharpe:1998xm,Rupak:2002sm,Bar:2003mh,Beane:2003xv,Tiburzi:2005vy}. These formulae typically involve low-energy constants (LECs) parameterizing the physics of the QCD scale (and shorter) whose values are unknown.  These LECs and corresponding operators (as well as quantum loop contributions) associated with lattice spacing effects, represent unphysical lattice spacing artifacts which must be removed from the correlation functions to obtain results in the continuum.  The remaining LECs, which survive the continuum limit, represent the short-distance QCD physics of interest.  One can view the role of the lattice calculations as determining the value of these physical constants, which when combined with the quantum loops, completely describe the low energy dynamics of QCD.

The energy of a system can be obtained from the long time behavior of Euclidean correlation functions. In practice, baryonic observables  (specially those connected to excited   or multi-baryon states), suffer from a signal-to-noise degradation at large times~\cite{lepage_error}%
\footnote{Recently, a ``restless pions" boundary condition has been proposed~\cite{Bedaque:2007pe} designed to improve this signal-to-noise problem.  Combined with anisotropic lattices, this may allow for a significant increase in the available time-slices to study the ground-state of these heavy systems.} 
while at short times, contamination from the many excited states masks the information about the low lying states.  This naturally leads to the use of anisotropic lattice actions~\cite{Karsch:1982ve,Burgers:1987mb}, in which the lattice spacing in time is taken to be $a_t =  a_s/\xi$.  Anisotropic lattices have been used extensively in the study of heavy quarks and quarkonia~\cite{Alford:1996nx,Klassen:1998ua}, glueballs~\cite{Morningstar:1997ff,Morningstar:1999rf}, and excited state baryon spectroscopy~\cite{Basak:2006ww}.  It is also anticipated that anisotropic lattices may aid in the study of nucleon-nucleon~\cite{Fukugita:1994ve,Beane:2006mx} and hyperon-nucleon~\cite{Beane:2006gf} interactions on the lattice.  For heavy systems made of light quarks, such as nucleon-nucleon systems, 
the rapid degradation of the signal-to-noise ratio of the correlation functions is particularly problematic in identifying the ground state.  Thus, when the excited state contamination has finally died away, there are very few time slices left before the noise dominates the correlation functions, leaving only a very narrow window in time to study these systems.  With the use of anisotropic actions, one may gain more useful information allowing an earlier identification of the ground state plateau.%

The production of a large ensemble of dynamical anisotropic Wilson gauge configurations is currently underway~\cite{Edwards:AllHands2007}.  It is therefore the aim of this work to construct the low energy effective theory which encodes these new anisotropic effects, so that they might be studied and systematically removed from the information extracted from the correlation functions generated with these lattices.  We begin by constructing the Symanzik action~\cite{Symanzik:1983dc,Symanzik:1983gh} for both the Wilson~\cite{Wilson:1974sk} and $\mathcal{O}(a)$ improved Wilson~\cite{Sheikholeslami:1985ij} actions in Sec.~\ref{sec:Ani_Wilson_action}.  Then in Sec.~\ref{sec:aniso_EFT}, we construct the chiral Lagrangians~\cite{Weinberg:1978kz,Gasser:1983yg,Gasser:1984gg} relevant for these anisotropic lattices for both mesons and baryons, focussing on the new effects arising from the anisotropy.  We also provide extrapolation formulae for these hadrons with their modified dispersion relations.  In Sec.~\ref{sec:Aoki_regime}, we highlight an important feature of anisotropic actions; for a fixed spatial lattice spacing and bare fermion mass, if the isotropic action is in the QCD-phase, this does not guarantee the anisotropic action is outside the Aoki-phase~\cite{Aoki:1983qi}.  We then conclude in Sec.~\ref{sec:concl}.

%
%
\section{Anisotropic Wilson lattice Action}\label{sec:Ani_Wilson_action}

The starting point for our discussion is the anisotropic lattice action and its symmetries, from which we will construct the continuum effective Symanzik action~\cite{Symanzik:1983dc,Symanzik:1983gh}, which will then allow us to construct the low-energy EFT describing the hadronic interactions including the dominant lattice spacing artifacts~\cite{Sharpe:1998xm}.  For the isotropic Wilson (and $\mc{O}(a)$ improved~\cite{Sheikholeslami:1985ij} Wilson) action, this program has been carried out to $\mc{O}(a^2)$ for the mesons~\cite{Sharpe:1998xm,Rupak:2002sm,Bar:2003mh} and baryons~\cite{Beane:2003xv,Tiburzi:2005vy}.  This work is a generalization of the previous work, extending the low-energy Wilson EFT to include the dominant lattice artifacts associated with the anisotropy.

The $\mc{O}(a)$-improved anisotropic lattice action, in terms of dimensionless fields, is given by~\cite{Klassen:1998fh,Chen:2000ej}
\begin{align}
S^\xi =&S_1^\xi+S_2^\xi,\\
S_1^\xi=& \beta \sum_{t,i<j} \frac{1}{\xi_0} P_{ij}(U) 
		+\beta \sum_{t,i} \xi_0 P_{ti}(U) 
	+\sum_{n} \bar{\psi}_n \bigg[ a_t m_0 + W_t (U) +\frac{\nu}{\xi_0} W_i(U) \bigg] \psi_n,\\
S_2^\xi=&	-\bar{\psi}_n \bigg[ c_t\, \sigma_{ti} \hat{F}_{ti}(U) 
		+\sum_{i<j} \frac{c_r}{\xi_0}\, \sigma_{ij} \hat{F}_{ij}(U) \bigg] \psi_n\, .
\end{align}
Here, $\xi_0$ is the bare anisotropy, $P_{ij}$ and $P_{ti}$ are space-space and space-time plaquettes of the gauge links $U$.  The bare (dimensionless) quark mass is $a_t m_0$, $W_t(U)$ and $W_i(U)$ are Wilson lattice derivatives and $\nu$ is a parameter which must be tuned to correct the ``speed of light."  The fields $\hat{F}_{ti}(U)$ and $\hat{F}_{ij}(U)$ are lattice equivalents of the gauge field-strength tensor in the space-time and space-space directions.  $S_1^\xi$ is the unimproved
Wilson action and $S_2^\xi$ is the anisotropic generalization of the Sheikholeslami-Wohlert term. The coefficients, $c_t$ and $c_r$, appearing in $S_2^\xi$ are needed for $\mc{O}(a)$ improvement of the anisotropic Wilson lattice action~\cite{Sheikholeslami:1985ij}.  At the classical level, they have been determined to be~\cite{Chen:2000ej}
\begin{align}\label{eq:ct_cr}
	c_t &= \frac{1}{2}\left( \frac{1}{\xi} + \nu \right)\, ,
	\nonumber\\
	c_r &= \nu\, ,
\end{align}
where $\xi = a_s / a_t$ is the renormalized anisotropy.%
\footnote{These parameters can also be tadpole improved with no more effort than in the isotropic action~\cite{Liao:2001yh}.} 
The choice of using $\xi$ as opposed to $\xi_0$ is conventional and the difference amounts to a slightly different value of $\nu$.

This anisotropic lattice theory retains all the symmetries of the Wilson action, except for the hypercubic invariance, and thus respects parity, time-reversal, translational invariance, charge-conjugation and cubic invariance.  In addition, for suitably tuned bare fermion masses, $a_t m_0,$ the theory has an approximate chiral symmetry, $SU(N_f)_L \otimes SU(N_f)_R$, which spontaneously breaks to the vector subgroup.%
\footnote{As mentioned in the Introduction, in Sec.~\ref{sec:Aoki_regime}, we will discuss Aoki-phases on anisotropic lattices which are not forbidden even if the equivalent isotropic action is not in the Aoki regime.} 
%

%
%
\subsection{Anisotropic Symanzik Action \label{sec:quarks}}

We begin by constructing the Symanzik Lagrangian for the {\it unimproved} anisotropic Wilson $S_1^\xi$ lattice action.  This will allow us to set our conventions and introduce a new basis of improvement terms which is advantageous to studying the new lattice artifacts which are remnants of the anisotropy.  In terms of dimensionful fields, the anisotropic Symanzik action is given by
\begin{align}
	S_{Sym}^\xi &=\int d^4x  \mc{L}_{Sym}^\xi\, ,
	\nonumber\\
	\mc{L}_{Sym}^\xi &= \mc{L}_{Sym}^{\xi(4)} + a_s \mc{L}_{Sym}^{\xi(5)}
		+ a_s^2 \mc{L}_{Sym}^{\xi(6)}\, ,
\end{align}
where we have conventionally chosen to use the spatial lattice spacing as our Symanzik expansion parameter.  In terms of the dimensionful fermion fields
\begin{equation}
	q \sim \frac{1}{a_s^3} \psi \quad,\quad 
	\bar{q} \sim \frac{1}{a_s^3} \bar{\psi} \, ,
\end{equation}
the anisotropic Lagrangian is given through $\mc{O}(a)$ by
\begin{align}\label{eq:sym_a0}
\mc{L}_{Sym}^{\xi} &= \bar{q} \left[ \Dslash +m_q \right] q
	+a_s\, \bar{q} \Big[ \bar c_t\, \sigma_{ti} F_{ti} +\bar c_r\, \sum_{i<j} \sigma_{ij} F_{ij} \Big] q\, .
\end{align}
 We have assumed that the parameter $\nu$ has been tuned in such a way as to make the breaking of $O(4)$ symmetry to vanish in the continuum limit. Otherwise, the quark kinetic term would separate into two terms, with an additional free parameter appearing in eq.~(\ref{eq:sym_a0}).  To clearly identify the new lattice spacing effects associated with the anisotropy, it is useful to work with the basis,
\begin{align}
\mc{L}_{Sym}^{\xi} &= \bar{q}\, \left[ \Dslash +m_q \right]\, q
	+a_s \frac{\bar c_r}{2}\, \bar{q}\, \sigma_{\mu\nu} F_{\mu\nu}\, q
	+a_s(\bar c_t-\bar c_r)\, \bar{q}\, \sigma_{ti} F_{ti}\, q\, ,
\end{align}
from which we recognize the first $\mc{O}(a)$ term as the Sheikholeslami-Wohlert~\cite{Sheikholeslami:1985ij} term which survives the isotropic limit, $c_{SW} = \bar c_r / 2$.  The second $\mc{O}(a)$ term $c_{SW}^\xi = (\bar c_t-\bar c_r)$, is an artifact of the anisotropy and the focus of this work. 
We can classify the effects from this operator and subsequent anisotropy operators at higher orders into two categories: those which contribute to physical quantities in a fashion similar to the lattice spacing artifacts already present, and those which introduce new hypercubic breaking effects.  The first type of effect will be difficult to distinguish from the already present lattice spacing artifacts which survive the isotropic limit.%
\footnote{Assuming a given set of lattice calculations are close to the continuum limit, and that a range of anisotropies is employed, one can disentangle the lattice artifacts associated with the new anisotropic operator from those which survive the isotropic limit.} 
The second category of effects are unique to anisotropic actions, and therefore more readily identifiable from correlation functions.

A useful manner to quantify these new anisotropic effects is to recognize that the anisotropy introduces a direction into the theory, which we can denote with the four-vector
\begin{equation}\label{eq:u_aniso}
	u^\xi_\mu = (1, \mathbf{0})\, .
\end{equation}
 This allows us to re-write the anisotropic Lagrangian in a manner amenable to spurion analysis
\begin{align}\label{eq:L_W^xi}
\mc{L}_W^{\xi} &	= \bar{q} \left[ \Dslash +m_q \right] q
	+a_s\, \bar{q}\, \Big[ c_{SW}\, \sigma_{\mu\nu} F_{\mu\nu}
		+c_{SW}^{\xi} u^\xi_\mu u^\xi_\nu\, \sigma_{\mu\lambda} F_{\nu\lambda} 
	\Big] q\, .
\end{align}
We then promote both $a_s c_{SW}$ and $a_s c_{SW}^\xi u^\xi_\mu u^\xi_\nu$ to spurion fields, transforming under chiral transformations in such a way as to conserve chiral symmetry,
\begin{align}
	&a_s c_{SW} \longrightarrow L (a_s c_{SW}) R^\dagger \ ,&
	& (a_s c_{SW})^\dagger \longrightarrow R (a_s c_{SW})^\dagger L^\dagger &
	\\
	&a_s c_{SW}^\xi u^\xi_\mu u^\xi_\nu \longrightarrow 
		L (a_s c_{SW}^\xi u^\xi_\mu u^\xi_\nu) R^\dagger \ ,&
	&(a_s c_{SW}^\xi u^\xi_\mu u^\xi_\nu)^\dagger \longrightarrow 
		R (a_s c_{SW}^\xi u^\xi_\mu u^\xi_\nu)^\dagger L^\dagger &
\end{align}
By constraining $a_s c_{SW}$ and $a_s c_{SW}^\xi u^\xi_\mu u^\xi_\nu$ and their hermitian conjugates to be proportional the flavor identity, they both explicitly break the $SU(N_F)_L \otimes SU(N_F)_R$ chiral symmetry down to the vector subgroup, just as the quark mass term.  In addition, we promote $a_s c_{SW}^\xi u^\xi_\mu u^\xi_\nu$ to transform under hypercubic transformations, so as to conserve hypercubic symmetry,
\begin{equation}
	a_s c_{SW}^\xi u^\xi_\mu u^\xi_\nu \longrightarrow 
		a_s c_{SW}^\xi u^\xi_\rho u^\xi_\sigma\, \Lambda_{\mu \rho} \Lambda_{\nu \sigma}\, .
\end{equation}
By constraining $u^\xi_\mu = (1, \mathbf{0})$, this spurion explicitly breaks the hypercubic symmetry of the action down to the cubic sub-group.  

It is worth noting that, in fact, at $\mc{O}(a)$, the anisotropic Symanzik action retains an accidental $O(3)$ symmetry in the spatial directions.  Close to the continuum, isotropic lattice actions retain an accidental Euclidean $O(4)$ (Lorentz) symmetry as the operators required to break this symmetry are of higher dimension and thus become irrelevant in the continuum limit~\cite{Symanzik:1983dc,Symanzik:1983gh}.  This phenomena is observed in the isotropic limit where the $O(4)$ symmetry is broken by the operator $a^2 \bar{q}\, \g_\mu D_\mu D_\mu D_\mu\, q$.   For unimproved anisotropic Wilson fermions, the breaking of the the hypercubic to cubic symmetry (which can be viewed as the breaking of the accidental $O(4)$ to the accidental $O(3)$ symmetry) occurs one order lower in the lattice spacing, at $\mc{O}(a)$, and therefore this will likely be a larger lattice artifact than the $O(4)$ breaking of the isotropic action.  We now perform a similar analysis for the $\mc{O}(a^2)$ Symanzik action.

%
%
\subsubsection{$\mc{O}(a^2)$ Symanzik Lagrangian}
In Ref.~\cite{Bar:2003mh}, the complete set of $\mc{O}(a^2)$ operators in the isotropic Symanzik action for Wilson fermions was enumerated, including the quark bi-linears and four-quark operators.  From an EFT point of view, it is useful to classify these operators in three categories, those operators which do not break any of the continuum (approximate) symmetries, those which explicitly break chiral symmetry and those which break Lorentz symmetry.  Most of the $\mc{O}(a^2)$ operators belong to the first category.  Because of their nature, they are the most difficult to determine and ultimately lead to a polynomial dependence in the lattice spacing of all correlation functions computed on the lattice (which can be parameterized as a polynomial dependence in $a$ of all the coefficients of the chiral Lagrangian).  The second set of operators, those which explicitly break chiral symmetry, can be usefully parameterized within an EFT framework as is commonly done with chiral Lagrangians extended to include lattice spacing artifacts~\cite{Sharpe:1998xm,Rupak:2002sm,Bar:2003mh,Beane:2003xv,Tiburzi:2005vy}.  The last set of operators which break Lorentz symmetry, can also be usefully studied in an EFT framework.  In the meson Lagrangian, these effects are expected to be small as they do not appear until $\mc{O}(p^4 a^2)$~\cite{Bar:2003mh}, while in the heavy baryon Lagrangian, these effects appear  at $\mc{O}(a^2)$~\cite{Tiburzi:2005vy}.  To distinguish these Lorentz breaking terms from the general lattice spacing artifacts appearing at $\mc{O}(a^2)$, one must study the dispersion relations of the hadrons, and not merely their ground states.  This is also generally true of all the anisotropic lattice artifacts which we now address.

In the construction of the anisotropic action, it is also beneficial to categorize the operators into several categories along the lines of those in the isotropic action mentioned above.  We do not show all of the new operators, as their explicit form will not be needed, but instead provide a representative set of the new anisotropic operators which illustrate the new lattice-spacing artifacts.  In the first category, we begin with operators which in the isotropic limit do not break any of the continuum symmetries.  Using the notation of Ref.~\cite{Bar:2003mh}, and using a superscript-${}^\xi$ to denote the new operators due to the anisotropy, we find for example 
\begin{align}
&O_3^{(6)} \longrightarrow \{ O_3^{(6)}\ ,\ {}^\xi O_3^{(6)} \}
	= \{ \bar{q}\, D_\mu \Dslash D_\mu \,q\ ,\ \bar{q}\, D_t \Dslash D_t\, q \}&
	\nonumber\\
&O_{11}^{(6)} \longrightarrow \{ O_{11}^{(6)}\ ,\  {}^\xi O_{11}^{(6)} \}
	= \{ ( \bar{q}\, \g_\mu q ) (\bar{q}\, \g_\mu q )\ ,\ ( \bar{q}\, \g_t q ) (\bar{q}\, \g_t q ) \} &\, .
\end{align}
In the second category, operators which explicitly break chiral symmetry, we find
\begin{equation}
O_{13}^{(6)} \longrightarrow \{ O_{13}^{(6)}\ ,\ {}^\xi O_{13}^{(6)} \}
	= \{ ( \bar{q}\, \s_{\mu\nu} q ) (\bar{q}\, \s_{\mu\nu} q )\ ,\  ( \bar{q}\, \s_{ti} q ) (\bar{q}\, \s_{ti} q )\} \, ,
\end{equation}
from which we observe that there is an operator which both breaks chiral and hypercubic symmetry.  The last category of operators stems from the Lorentz breaking operator in the isotropic limit,
\begin{equation}
O_4^{(6)} \longrightarrow \{ O_4^{(6)}\ ,\ {}^\xi O_4^{(6)} \}
	=\{ \bar{q}\, \g_{\mu} D_\mu D_\mu D_\mu\, q\ ,\  \bar{q}\, \g_{i} D_i D_i D_i\, q\} \, ,
\end{equation}
from which we note that there is only one operator which breaks the accidental $O(3)$ symmetry down to the cubic group, ${}^\xi O_4^{(6)}$, and therefore the dominant $O(3)$ breaking artifacts in principle can be completely removed from the theory by studying the dispersion relation of only one hadron, for example the pion.  Each of these new operators can be written in their spuriously hypercubic-invariant form by making use of the anisotropic  vector we introduced in Eq.~\eqref{eq:u_aniso},
\begin{align}\label{eq:asq_aniso}
	&{}^\xi O_3^{(6)} = u^\xi_\mu u^\xi_\nu\, \bar{q}\, D_\mu \Dslash D_\nu\, q,&
	\nonumber\\
	&{}^\xi O_{11}^{(6)} = u^\xi_\mu u^\xi_\nu\, ( \bar{q}\, \g_\mu q ) (\bar{q}\, \g_\nu q ),&
	\nonumber\\
	&{}^\xi O_{13}^{(6)} = u^\xi_\mu u^\xi_\nu\, (\bar{q}\, \s_{\mu\lambda} q) (\bar{q}\, \s_{\nu\lambda} q),&
	\nonumber\\
	&{}^\xi O_4^{(6)} = 
		\bar{\d}^\xi_{\mu\mu^\prime}
		\bar{\d}^\xi_{\mu\nu^\prime}
		\bar{\d}^\xi_{\mu\rho^\prime}
		\bar{\d}^\xi_{\mu\sigma^\prime}
		\bar{q}\, \g_{\mu'} D_{\nu'} D_{\s'} D_{\rho'}\, q,& 
\end{align}
and similarly for the rest of the dimension-6 anisotropic operators, ${}^\xi O_{1-8}^{(6)}$, and ${}^\xi O_{11-18}^{(6)}$.  In this equation, for we have defined
\begin{equation}
	\bar{\d}^\xi_{\mu\nu} \equiv \delta_{\mu\nu} - u^\xi_\mu u^\xi_\nu\, .
\end{equation}
Most of these operators do not break chiral symmetry, and therefore are present for chirally symmetric fermions such as domain-wall~\cite{Kaplan:1992bt,Shamir:1993zy,Furman:1994ky} and overlap~\cite{Narayanan:1992wx,Narayanan:1994gw,Neuberger:1997fp} fermions.  We now proceed to construct the anisotropic chiral Lagrangian.

%
%
\section{Anisotropic Chiral Lagrangian \label{sec:aniso_EFT}}

Now that we have the complete set of Symanzik operators through $\mc{O}(a^2)$ relevant to the anisotropic Wilson action and the $\mc{O}(a)$ improved version thereof, we can construct the equivalent operators in the chiral Lagrangian which encode these new anisotropic artifacts.  We begin with the meson Lagrangian and then move to the heavy baryon Lagrangian.  

%
%
\subsection{Meson Chiral Lagrangian \label{sec:mesons}}

We construct the chiral Lagrangian using a spurion analysis of the quark level Lagrangian given in Eqs.~\eqref{eq:L_W^xi} and \eqref{eq:asq_aniso}.  We generally assume a power counting
\begin{equation}
	m_q \sim a \Lambda^2\, ,
\end{equation}
but work to the leading order necessary to parameterize the dominant artifacts from the anisotropy.  At LO, the meson Lagrangian is given by%
\footnote{We remind the reader we are working in Euclidean spacetime.  We are using the convention $f\sim 132$~MeV.} 
\begin{align}\label{eq:MesonsLO}
\mc{L}_\phi^\xi =&\ 
	\frac{f^2}{8} \tr \left( \partial_\mu \Sigma \partial_\mu \Sigma^\dagger \right)
	-\frac{f^2}{4} \tr \left( m_q\mathbb{B} \Sigma^\dagger + \Sigma (m_q\mathbb{B})^\dagger \right) 
	\nonumber\\&
	-\frac{f^2}{4} \tr \left( a_s \mathbb{W} \Sigma^\dagger + \Sigma (a_s \mathbb{W})^\dagger \right)
	-\frac{f^2}{4} \tr \left( a_s \mathbb{W}^\xi \Sigma^\dagger + \Sigma (a_s \mathbb{W}^\xi)^\dagger \right)\, .
\end{align}
By taking functional derivatives with respect to the spurion fields in both the quark level and chiral level actions, one can show~\cite{GellMann:1968rz}
\begin{equation}
	\mathbb{B} = \lim_{m_q\rightarrow 0} \frac{ | \langle \bar{q} q \rangle |}{f^2}\, ,
\end{equation}
and similarly the new dimension-full chiral symmetry breaking parameters are defined as,
\begin{align}
	\mathbb{W} &= \lim_{m_q \rightarrow 0}\ 
		c_{SW} \frac{ \langle \bar{q} \sigma_{\mu\nu} F_{\mu\nu} q \rangle}{f^2}\, ,
	\\
	\mathbb{W}^\xi &= \lim_{m_q \rightarrow 0}\ 
		c_{SW}^\xi \frac{ \langle \bar{q} \sigma_{ti} F_{ti} q \rangle}{f^2}
	\nonumber\\
		&= \lim_{m_q \rightarrow 0}\ 
		c_{SW}^\xi u^\xi_\mu u^\xi_\nu 
		\frac{ \langle \bar{q} \sigma_{\mu\lambda} F_{\nu\lambda} q \rangle}{f^2}\, .
\end{align}
This anisotropic Lagrangian, Eq.~\eqref{eq:MesonsLO}, has one more operator than in the isotropic limit~\cite{Sharpe:1998xm,Rupak:2002sm}, which is simply a reflection that there are now two distinct $\mc{O}(a)$ operators in the Symanzik action.  However, for a fixed anisotropy, $\xi = a_s /a_t$, these two $\mc{O}(a)$ operators are indistinguishable and can be combined into one.  Inserting the tree level values of the coefficients in the Symanzik action~\cite{Chen:2000ej}, one finds
\begin{align}
	\mathbb{W} &\propto c_{SW} = \nu\, ,
	\\
	\mathbb{W}^\xi &\propto c_{SW}^\xi = \frac{1}{2} \left( \frac{a_s}{a_t} -\nu \right)\, ,
\end{align}
where the speed-of-light parameter, $\nu$, is determined in the tuning of the anisotropic action.  The first clear signal of the anisotropy begins at the next order, $\mc{O}(a p^2)$ for the unimproved action and $\mc{O}(a^2 p^2)$ for the improved action.  We first discuss the unimproved case.

These next set of operators are only present for the unimproved action.  In the isotropic limit, it was shown there are five additional operators at this order~\cite{Rupak:2002sm}
\begin{align}\label{eq:Wilson_Oam}
\mc{L}_{\phi,am} =&\ 2 W_4 \tr \left( \partial_\mu \S \partial_\mu \S^\dagger \right)
		\tr \left( a_s \mathbb{W} \Sigma^\dagger + \Sigma (a_s \mathbb{W})^\dagger \right)
	\nonumber\\& 
	+2W_5 \tr \left( \partial_\mu \S \partial_\mu \S^\dagger 
		\left[ a_s \mathbb{W} \Sigma^\dagger + \Sigma (a_s \mathbb{W})^\dagger \right] \right)
	\nonumber\\& 
	+4 W_6 \tr \left( m_q\mathbb{B} \Sigma^\dagger + \Sigma (m_q\mathbb{B})^\dagger \right)
		\tr \left( a_s \mathbb{W} \Sigma^\dagger + \Sigma (a_s \mathbb{W})^\dagger \right)
	\nonumber\\&Ä
	+4 W_7 \tr \left( m_q\mathbb{B} \Sigma^\dagger - \Sigma (m_q\mathbb{B})^\dagger \right)
		\tr \left( a_s \mathbb{W} \Sigma^\dagger - \Sigma (a_s \mathbb{W})^\dagger \right)
	\nonumber\\&
	+4 W_8 \tr \left( m_q\mathbb{B} \S^\dagger a_s \mathbb{W} \S^\dagger
		+\S (m_q \mathbb{B})^\dagger \S (a_s \mathbb{W})^\dagger \right)\, .
\end{align}
For the anisotropic action, there are an additional five operators similar to the above five with the replacement of low-energy constants (LECs) $W_{4-8} \rightarrow W_{4-8}^\xi$ and a simultaneous replacement of the condensate $\mathbb{W} \rightarrow \mathbb{W}^\xi$.  These five new operators, as with the $\mc{O}(a)$ operators, are indistinguishable from those in Eq.~\eqref{eq:Wilson_Oam} at a fixed anisotropy.  There are two additional operators at this order however, which introduce new effects associated with the anisotropy,
\begin{align}
\mc{L}_{\phi,am}^{\xi} =&\ W_1^\xi \tr \left( \partial_\mu \S \partial_\nu \S^\dagger \right)
		\tr \left( u^\xi_\mu u^\xi_\nu \left[ 
			a_s \mathbb{W}^\xi \Sigma^\dagger + \Sigma (a_s \mathbb{W}^\xi)^\dagger \right] \right)
	\nonumber\\& 
	+W_2^\xi \tr \left( \partial_\mu \S \partial_\nu \S^\dagger\, 
		u^\xi_\mu u^\xi_\nu \left[ 
			a_s \mathbb{W}^\xi \Sigma^\dagger + \Sigma (a_s \mathbb{W}^\xi)^\dagger \right] \right)\, .
\end{align}
When the anisotropic  vectors are set to their constant value, $(u^\xi_\mu)^T = (1,\mathbf{0})$, one sees that these operators lead to a modification of the pion (meson) dispersion relation,
\begin{multline}\label{eq:pion_dispersion_Oa}
	(E_\pi^2+p_\pi^2)\left( 1 +W  \frac{a_s \mathbb{W}}{f_\pi^2} \right)  \longrightarrow 
	\\
	(E_\pi^2+p_\pi^2) \left( 1 
		+W \frac{a_s\mathbb{W}}{f_\pi^2} +W^\xi  \frac{a_s\mathbb{W}^\xi}{f_\pi^2} \right)
	+E_\pi^2\, \tilde{W}^\xi  \frac{a_s \mathbb{W}^\xi}{f_\pi^2}\, ,
\end{multline}
where 
\begin{align}
	&W = 32(N_f W_4 +W_5)\, ,&
	&W^\xi = 32(N_f W_4^\xi +W_5^\xi)\, ,&
	\nonumber\\
	&\tilde{W}^\xi = 16(N_f W_1^\xi + W_2^\xi)\, ,
\end{align}
and $N_f$ is the number of fermion flavors.  With the $\mc{O}(a)$ improved anisotropic action, these effects all vanish and the leading lattice artifacts begin at $\mc{O}(a^2)$.  The chiral Lagrangian at this next order in the isotropic limit was determined in Ref.~\cite{Bar:2003mh} for which there were three new operators.  In the anisotropic theory, there are an additional six operators but just as with the $\mc{O}(a)$ Lagrangian, the three new anisotropic operators can not be distinguished from those which survive the isotropic limit unless multiple values of the anisotropy are used.  The Lagrangian at this order is
\begin{align}
\mc{L}_{\phi,a^2}^{\xi} =& -4W_6^\prime\, 
		\Big[ a_s \mathbb{W}\, \tr \left( \S +\S^\dagger \right) \Big]^2
	-4\hat{W}_6^\xi\,  
		\Big[ a_s \mathbb{W}^\xi\, \tr \left( \S +\S^\dagger \right) \Big]^2
	\nonumber\\&
	-4W_7^\prime\, 
		\Big[ a_s \mathbb{W}\, \tr \left( \S -\S^\dagger \right) \Big]^2
	-4\hat{W}_7^\xi\,  
		\Big[ a_s \mathbb{W}^\xi\, \tr \left( \S -\S^\dagger \right) \Big]^2
	\nonumber\\&
	-4W_8^\prime\, (a_s \mathbb{W})^2\, 
		\tr \left( \S\S +\S^\dagger \S^\dagger \right)
	-4\hat{W}_8^\xi\, (a_s \mathbb{W}^\xi)^2\, 
		\tr \left( \S\S +\S^\dagger \S^\dagger \right)
	\nonumber\\&
	-4\bar{W}_6^\xi\, (a_s \mathbb{W})(a_s \mathbb{W}^\xi)
		\Big[ \tr \left( \S +\S^\dagger \right) \Big]^2
	-4\bar{W}_7^\xi\, (a_s \mathbb{W})(a_s \mathbb{W}^\xi)
		\Big[ \tr \left( \S -\S^\dagger \right) \Big]^2
	\nonumber\\&
	-4\bar{W}_8^\xi\, (a_s \mathbb{W})(a_s \mathbb{W}^\xi)\, 
		\tr \left( \S\S +\S^\dagger \S^\dagger \right)\, .
\end{align}
The last three operators in this Lagrangian vanish for an $\mc{O}(a)$-improved action as they are directly proportional to the product of the two $\mc{O}(a)$ terms in the Symanzik Lagrangian, Eq.~\eqref{eq:L_W^xi}.  The other six operators in this Lagrangian receive contributions both from products of the $\mc{O}(a)$ Symanzik operators as well as from terms in the $\mc{O}(a^2)$ Symanzik Lagrangian.%
\footnote{For conventional reasons~\cite{Bar:2003mh} we have normalized these operators to the square of the condensates appearing at $\mc{O}(a)$, but one should not confuse this to mean that these operators vanish for the $\mc{O}(a)$-improved action.} 
Therefore they are still present for an $\mc{O}(a)$-improved action but the numerical values of their LECs, $W_{6-8}^\prime$ and $\hat{W}_{6-8}^\xi,$ will be different in the improved case.   

At $\mc{O}(a^2 p^2)$, there are nine new operators, three of which survive the isotropic limit, and three of which explicitly break the accidental $O(4)$ symmetry down to $O(3)$,
\begin{align}\label{eq:asq_psq}
\mc{L}_{\phi,a^2p^2}^{\xi} =&\
	Q_1 (a_s \mathbb{W})^2 \tr \left( \partial_\mu \S \partial_\mu \S^\dagger \right)
	+Q_2 (a_s \mathbb{W})^2 
		\tr \left( \partial_\mu \S \partial_\mu \S^\dagger \right) \tr \left( \S +\S^\dagger \right)
	\nonumber\\& 
	+Q_3 (a_s \mathbb{W})^2 
		\tr \left( \partial_\mu \S \partial_\mu \S^\dagger \left[ \S +\S^\dagger \right] \right)
	+Q_1^\xi (a_s \mathbb{W}^\xi)^2 \tr \left( \partial_\mu \S \partial_\mu \S^\dagger \right)
	\nonumber\\& 
	+Q_2^\xi (a_s \mathbb{W}^\xi)^2 
		\tr \left( \partial_\mu \S \partial_\mu \S^\dagger \right) \tr \left( \S +\S^\dagger \right)
	+Q_3^\xi (a_s \mathbb{W}^\xi)^2 
		 \tr \left( \partial_\mu \S \partial_\mu \S^\dagger \left[ \S +\S^\dagger \right] \right)
	 \nonumber\\& 
	+\hat{Q}_1^\xi (a_s \mathbb{W}^\xi)^2 u^\xi_\mu u^\xi_\nu\ 
		\tr \left( \partial_\mu \S \partial_\nu \S^\dagger \right)
	+\hat{Q}_2^\xi (a_s \mathbb{W}^\xi)^2 u^\xi_\mu u^\xi_\nu\ 
		\tr \left( \partial_\mu \S \partial_\nu \S^\dagger \right)
		\tr \left( \S +\S^\dagger \right)
	\nonumber\\&
	+\hat{Q}_3^\xi (a_s \mathbb{W}^\xi)^2 u^\xi_\mu u^\xi_\nu\ 
	 \tr \left( \partial_\mu \S \partial_\nu \S^\dagger \left[ \S +\S^\dagger \right] \right)\, .
\end{align}
The first operator in each set of three with coefficients $Q_1$, $Q_1^\xi$ and $\hat{Q}_1^\xi$ are modifications of the LO kinetic operator, while the remaining operators additionally break chiral symmetry.  The first operator in Eq.~\eqref{eq:asq_psq} is an example of the operators mentioned before which do not break any of the (approximate) lattice symmetries, and are therefore amount to polynomial renormalizations of the continuum LECs.  In this example, we see with the replacement
\begin{equation}
	f^2 \rightarrow f^2 + 8 Q_1 (a_s \mathbb{W})^2\, ,
\end{equation}
the entire effects from this $\mc{O}(a^2 p^2)$ operator are renormalized away to all orders in the EFT.  In the anisotropic theory, this also works for the operator with LEC $Q_1^\xi$, as this operator does not explicitly break $O(4)$ symmetry.  This provides a further example of effects which arise because of the anisotropy but contribute to the low-energy dynamics in an isotropic fashion.  The operators with explicit anisotropic vectors will modify the dispersion relation as in Eq.~\eqref{eq:pion_dispersion_Oa} but at $\mc{O}(a^2)$.  For the $\mc{O}(a)$-improved action, these operators will provide the dominant modification to the pseudo-Goldstone dispersion relations.

The last set of operators we wish to address for the meson Lagrangian are those which explicitly break the accidental $O(3)$ symmetry down to the cubic group.  There are two operators in the meson chiral Lagrangian but they stem from only one quark level operator at $\mc{O}(a^2)$ and therefore all the LO $O(3)$ breaking effects for all hadrons can be removed with the inclusion and tuning of one new operator in the action, ${}^\xi O_4^{(6)}$ from Eq.~\eqref{eq:asq_aniso}.  The $\mc{O}(3)$ breaking operators in the meson chiral Lagrangian are
\begin{align}
\mc{L}_{\phi,O(3)}^\xi =&\ C_1^\xi (a_s \mathbb{W}^\xi)^2 
		\bar{\d}^\xi_{\mu\mu^\prime}
		\bar{\d}^\xi_{\mu\nu}
		\bar{\d}^\xi_{\mu\rho}
		\bar{\d}^\xi_{\mu\sigma}\ 
		\tr \left( \partial_{\mu^\prime} \S \partial_\nu \S^\dagger \right)
		\tr \left( \partial_\rho \S \partial_\s \S^\dagger \right)
	\nonumber\\&
	+C_2^\xi (a_s \mathbb{W}^\xi)^2 
			\bar{\d}^\xi_{\mu\mu^\prime}
	\bar{\d}^\xi_{\mu\nu}
	\bar{\d}^\xi_{\mu\rho}
	\bar{\d}^\xi_{\mu\sigma}\ 
	\tr \left( \partial_{\mu^\prime} \S \partial_\nu \S^\dagger \partial_\rho \S \partial_\s \S^\dagger \right)\, ,
\end{align}

%
%
\subsubsection{Aoki Regime \label{sec:Aoki_regime}}

  Aoki first pointed out the possibility that at finite lattice spacing, lattice actions can undergo spontaneous symmetry breaking of flavor and parity in certain regions of phase space~\cite{Aoki:1983qi}.  In Ref.~\cite{Sharpe:1998xm}, Sharpe and Singleton addressed this possibility within an EFT framework by extending the meson chiral Lagrangian to include lattice spacing contributions.  We begin by summarizing the discussion of Sharpe and Singleton which will allow us to highlight the new effects that arise from the anisotropy.  For clarity of discussion, we consider the unimproved two-flavor theory in the isospin limit.  Following the notation of Ref.~\cite{Sharpe:1998xm}, the non-kinetic part of the chiral potential can be written
\begin{equation}\label{eq:chiral_potential}
	\mc{V}_\chi = -\frac{c_1}{4} \tr \left( \S +\S^\dagger \right)
		+\frac{c_2}{16} \left[ \tr \left( \S + \S^\dagger \right) \right]^2\, ,
\end{equation}
where $c_1$ and $c_2$ are functions of the quark mass $m_q$ and the lattice spacing $a$,
\begin{align}
	c_1 &\sim \L^4 \left( \frac{m_q}{\L} + a \L \right) +\dots\, ,
	\nonumber\\
	c_2 &\sim \L^4 \left( \frac{m_q^2}{\L^2} +m_q a +a^2 \L^2 \right) +\dots\, .
\end{align}
The ``$\dots$" denote higher order terms in the quark mass and lattice spacing and we are ignoring dimensionless numbers of $\mc{O}(1)$.  Assuming a power counting $m_q \sim a\L^2$, the contributions to the vacuum from the $c_2$ term are suppressed compared to the $c_1$ term, and the vacuum is in the continuum phase with the chiral condensate aligned with unity.  The Aoki-phase can occur when there is a fine-tuning of the quark mass and lattice spacing contributions to $c_1$ such that overall size of $c_1$ becomes comparable to $c_2$.  Parameterizing the $\S$-field as
\begin{equation}
	\S = A +i \mathbf{\tau} \cdot \mathbf{B}\, ,
\end{equation}
with the constraint $A^2 +\mathbf{B}^2 =1$, the potential, Eq.~\eqref{eq:chiral_potential} becomes
\begin{equation}
	\mc{V}_\chi = -c_1 A +c_2 A^2\, .
\end{equation}
If the minimum of the potential occurs for $-1 < A_{0} < 1$, then the vacuum
\begin{equation}
\S_0 = \langle\, \S\, \rangle 
	= A_0 +i \mathbf{\tau} \cdot \mathbf{B_0}\, ,
\end{equation} 
develops a non-zero value of $\mathbf{B}_0$, spontaneously breaking both parity and the remnant vector-chiral symmetry, $SU(2)_V \rightarrow U(1)$~\cite{Sharpe:1998xm}, giving rise to one massive pseudo-Goldstone pion and two massless Goldstone pions.  If the minimum of the potential occurs for $A_{min} \leq -1$ or $A_{min} \geq 1$ then the vacuum lies along (or opposite) the identitiy, $|A_0| =1, |\mathbf{B}_0|=0$ with the same symmetry breaking pattern as QCD.

For unimproved Wilson fermions in the isotropic limit, the leading contributions to $c_1$ are
\begin{equation}
	c_1 = f^2 \Big( m_q \mathbb{B} +a_s \mathbb{W} \Big)\, .
\end{equation}
In the anisotropic theory, there is an additional contribution to the LO potential, such that
\begin{align}
c_1 \rightarrow c_1^\xi &= 
	c_1 + f^2 a_s \mathbb{W}^\xi
	\nonumber\\&
	= f^2 \Big( m_q \mathbb{B} 
	+a_s \mathbb{W} 
	+a_s \mathbb{W}^\xi \Big)\, .
\end{align}
If the two terms which contribute to 
$\mathbb{W} \propto 2 \langle \bar{q} \s_{ti} F_{ti} q + \bar{q} \s_{ij} F_{ij} q \rangle$ are of opposite sign, then $\mathbb{W}^\xi \propto \langle \bar{q} \s_{ti} F_{ti} q \rangle$ may be the dominant lattice spacing contribution to $c_1$.  Therefore, the anisotropic theory may be in the Aoki-regime even when the isotropic limit of the theory (with the same $a_s$) is not, and \textit{vice versa}.  This same discussion holds for $\mc{O}(a)$-improved actions as well but with a different power counting. 

\bigskip
This analysis carries important consequences for anisotropic actions with domain-wall and overlap fermions as well.  These actions are generally tuned to lie between the first two-fingers of the Aoki-regime in the $m_0 - g^2$ plane~\cite{Antonio:2007tr}.  The optimal value of the bare fermion mass, which provides the least amount of residual chiral symmetry breaking, may be shifted in the anisotropic theory relative to the isotropic value, and furthermore the allowed values of the coupling for which there is a QCD-phase may shift as well.

%
%
\subsection{Heavy Baryon Lagrangian \label{sec:baryons}}

\begin{figure}[t]
\center
\begin{tabular}{cccc}
\includegraphics[width=0.18\textwidth]{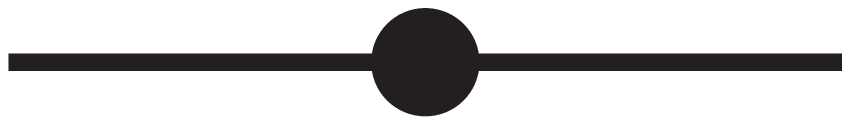} &
\includegraphics[width=0.18\textwidth]{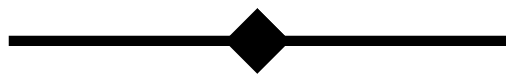} &
\includegraphics[width=0.26\textwidth]{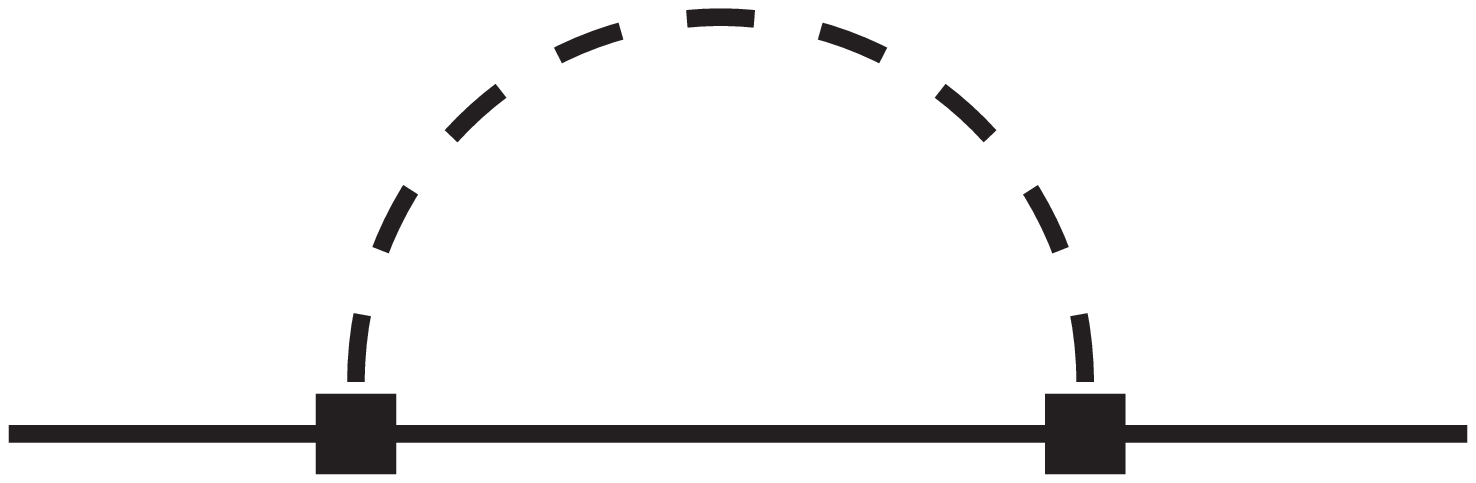} &
\includegraphics[width=0.26\textwidth]{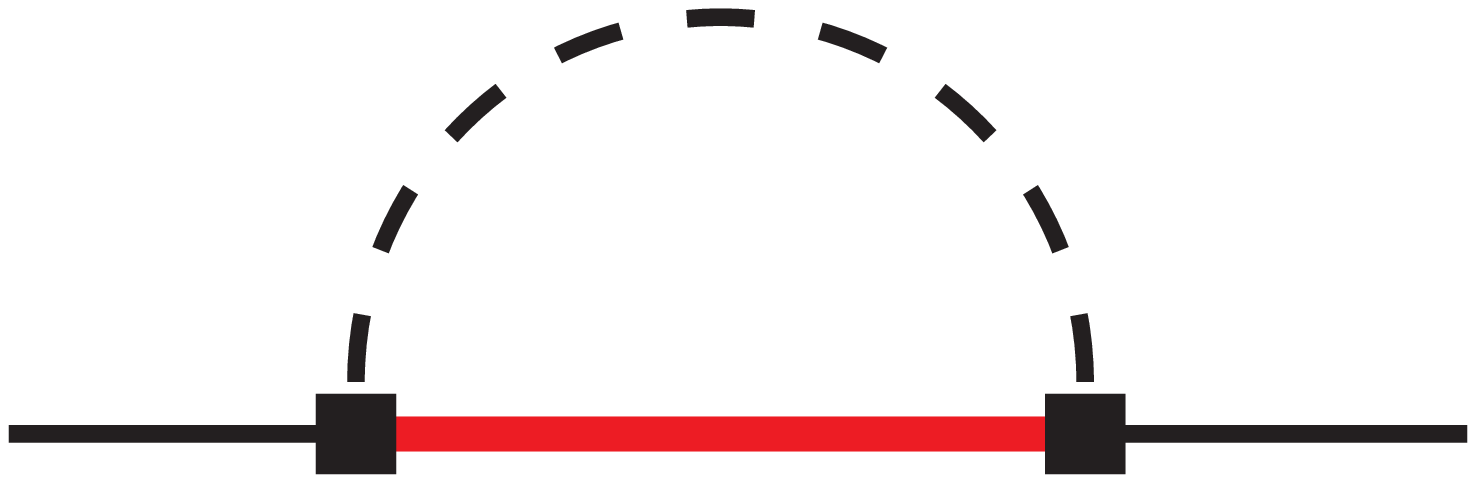} \\
(a) & (b) & (c) & (d)
\end{tabular}
\caption{\label{fig:NmassNLO} \textit{Diagrams contributing to the nucleon mass at LO ((a) and (b)) and NLO ((c) and (d)).  Figure (a) is an insertion of the leading quark mass term proportional to $\s_M$.  Figure (b) is an insertion of the lattice spacing terms proportional to $\s_W$ and $\s^\xi_W$.  The loop graphs arise from the pion-nucleon and pion-nucleon-delta couplings.  These loop graphs generically scale as $m_\pi^3$ but also depend upon $\D$, and away from the continuum limit upon the lattice spacing as well.  All vertices in these graphs are from the Lagrangian in Eq.~\eqref{eq:NTphiLag}.}}
\end{figure}

We now construct the operators in the baryon Lagrangian which encode the leading lattice artifacts from the anisotropy.  We use the heavy baryon formalism~\cite{Jenkins:1990jv,Jenkins:1991es} and explicitly construct the two-flavor theory in the isospin limit including nucleons, delta-resonances and pions.  The extension of this to include the octet and decuplet baryons is a straight forward exercise.  This construction builds upon previous work in which the heavy baryon Lagrangian has been extended in the isotropic limit to include the $\mc{O}(a)$~\cite{Beane:2003xv} and $\mc{O}(a^2)$~\cite{Tiburzi:2005vy} lattice artifacts for various baryon observables.%
\footnote{In those works~\cite{Beane:2003xv,Tiburzi:2005vy}, the heavy baryon Lagrangian was extended explicitly for Wilson fermions and in the latter also for a mixed action scheme with chirally symmetric valence fermions.  In this work, we explicitly address the anisotropic effects for these two scenarios.  One can also include the anisotropic effects for twisted mass~\cite{Frezzotti:2000nk} and staggered fermions~\cite{Kogut:1974ag,Susskind:1976jm} following the heavy baryon construction in Refs.~\cite{WalkerLoud:2005bt} and \cite{Bailey:2006zn,Bailey:2007iq} respectively.  However, we are unaware of specific plans to generate anisotropic twisted mass or staggered fermions so we do not pursue this here.} 
The Lagrangian is constructed as a perturbative expansion about the static limit of the baryon, treating it as a heavy, static source as with heavy quark effective theory~\cite{Georgi:1990um,Manohar:2000dt}.  In building the heavy baryon chiral Lagrangian it is useful to introduce a new chiral field,%
\footnote{This field is generally denoted as $\xi$ but to avoid confusion with the anisotropy parameter, we use $\s$.} 
\begin{equation}
	\s = \sqrt{\S} = \exp \left( \frac{i \phi}{f} \right)\, ,
\end{equation}
which transforms under chiral rotations as
\begin{equation}\label{eq:V}
	\s \rightarrow L \s V^\dagger = V \s R^\dagger\, ,
\end{equation}
with $V$ defined by Eq.~\eqref{eq:V}.  One then dresses the baryon fields with $\s$ such that under chiral transformations, the nucleon and delta fields transform as
\begin{align}
	N_i &\rightarrow V_{ij} N_j\, ,
	\nonumber\\
	T_{ijk} &\rightarrow V_{ii^\prime}V_{jj^\prime}V_{kk^\prime} T_{i^\prime j^\prime k^\prime}\, ,
\end{align}
where $N$ is a two-component flavor field and $T$ is a flavor-symmetric rank-three tensor, normalized such that
\begin{align}
	&&
	& N_1 = p,&
	& N_2 = n,& &&
	\nonumber\\
	&T_{111} = \D^{++},&
	&T_{112} = \frac{1}{\sqrt{3}} \D^{+},&
	&T_{122} = \frac{1}{\sqrt{3}} \D^{0},&
	&T_{222} = \D^{-}.&
\end{align}
Both the nucleon and delta are treated as heavy matter fields and are constrained by the relations
\begin{align}
	N &= \frac{1+\vslash}{2} N\, ,
	\nonumber\\
	T_\mu &= \frac{1 +\vslash}{2} T_\mu
\end{align}
where $v_\mu$ is the four velocity of the baryon which can be chosen in its rest frame to be $v_\mu = (1, \mathbf{0})$.  This has the effect of projecting onto the particle component of the field in the rest frame of the baryon.  The spin-3/2 fields can be described with a Rarita-Schwinger field, which in the heavy baryon formalism gives rise to the constraints
\begin{equation}
	v \cdot T = 0\quad, \quad S \cdot T = 0\, ,
\end{equation}
where $S_\mu$ is a covariant spin vector~\cite{Jenkins:1990jv,Jenkins:1991es}.  The simultaneous inclusion of the nucleon and delta fields introduces a new parameter into the Lagrangian, the delta-nucleon mass splitting in the chiral limit,
\begin{equation}\label{eq:Delta}
	\Delta = m_T - m_N \Big|_{m_q = 0}\, ,
\end{equation}
which is generally counted as $\D \sim m_\pi$ in the chiral power counting~\cite{Jenkins:1990jv,Jenkins:1991es,Hemmert:1997ye}.  Because this mass parameter is a chiral singlet, it leads to a modification of all the LECs in the heavy baryon Lagrangian.  These particular effects can be systematically accounted for be treating all LECs as polynomials in $\D$~\cite{WalkerLoud:2004hf,Tiburzi:2004rh,Tiburzi:2005na,WalkerLoud:2006sa}.  In the isospin limit, the heavy baryon Lagrangian, including the leading quark mass terms, the leading lattice spacing terms and the leading baryon-pion couplings is given by
\begin{align}\label{eq:NTphiLag}
	\mc{L}_{NT\phi} =&\ \bar{N} v \cdot D N 
		+a_t \sigma_{E} \bar{N} (v\cdot u^\xi u^\xi\cdot D) N
		-2 \sigma_M \bar{N} N \tr ( \mc{M}_+ ) 
	\nonumber\\&
		+ \bar{T}_\mu \big[ v \cdot D +\Delta \big] T_\mu
		+ a_t \bar{\sigma}_E \bar{T}_\mu (v\cdot u^\xi u^\xi\cdot D) T_\mu
		-2 \bar{\sigma}_M \bar{T}_\mu T_\mu \tr ( \mc{M}_+ ) 
	\nonumber\\&
		+2 g_A \bar{N} S \cdot \mc{A} N
		-2g_{\D\D} \bar{T}_\mu S \cdot \mc{A} T_\mu
		+g_{\D N} \big[ \bar{T}^{kji}_\mu \mc{A}_{i,\mu}^{\ i^\prime} \epsilon_{j i^\prime} N_k +h.c. \big]\, .
	\nonumber\\&
		- 2\sigma_W \bar{N} N \tr ( \mc{W}_+ )
		- 2\sigma_W^\xi \bar{N} N \tr ( \mc{W}^\xi_+ )
		-2 \bar{\sigma}_W \bar{T}_\mu T_\mu \tr ( \mc{W}_+ ) 
		-2 \bar{\sigma}_W^\xi \bar{T}_\mu T_\mu \tr ( \mc{W}^\xi_+ ) 
\end{align}
In this Lagrangian, the chiral symmetry breaking spurions are given by
\begin{align}
	\mc{M}_+ &= \frac{1}{2} \left( \s m_q^\dagger \s + \s^\dagger m_q \s^\dagger \right)\, ,
	\nonumber\\
	\mc{W}_+ &= \frac{a_s}{2} \left( \s \mathbb{W}^\dagger \s + \s^\dagger \mathbb{W} \s^\dagger \right)\, ,
	\nonumber\\
	\mc{W}^\xi_+ &= \frac{a_s}{2} \left( \s (\mathbb{W}^\xi)^\dagger \s 
		+ \s^\dagger \mathbb{W}^\xi \s^\dagger \right)\, ,
\end{align}
the chiral covariant derivative is
\begin{align}
	(D^\mu N)_i &= \partial^\mu N_i + \mc{V}^\mu_{ij}N_j\, ,
	\nonumber\\
	(D^\mu T)_{ijk} &= \partial^\mu T_{ijk} + \mc{V}^\mu_{i i^\prime} T_{i^\prime j k}
		+ \mc{V}^\mu_{j j^\prime} T_{ij^\prime k}
		+ \mc{V}^\mu_{k k^\prime} T_{ijk^\prime}
\end{align} 
and the vector and axial fields are respectively given by
\begin{align}
	\mc{V}_\mu &= \frac{1}{2} \left( \s \partial_\mu \s^\dagger + \s^\dagger \partial_\mu \s \right)\, ,
	\nonumber\\
	\mc{A}_\mu &= \frac{i}{2} \left( \s \partial_\mu \s^\dagger - \s^\dagger \partial_\mu \s \right)\, .
\end{align}
Relative to the Wilson extension of the heavy baryon Lagrangian in the isotropic limit~\cite{Beane:2003xv,Tiburzi:2005vy} there is one additional mass operator for the nucleon and delta fields.  There are additionally extra derivative operators which give rise to the leading modification of the dispersion relation for the baryon fields.  As in the meson chiral Lagrangian, Eq.~\eqref{eq:MesonsLO}, at fixed anisotropy, these two additional mass operators (with coefficients $\s_W^\xi$ and $\bar{\s}_W^\xi$) can not be distinguished from their counterparts which survive the isotropic limit (with coefficients $\s_W$ and $\bar{\s}_W$).  At this order in the lattice spacing, $\mc{O}(a)$, the baryon-pion couplings, $g_A$, $g_{\D N}$ and $g_{\D\D}$, are not modified~\cite{Beane:2003xv}.  This Lagrangian, Eq.~\eqref{eq:NTphiLag} gives rise to the LO and NLO mass corrections to the baryon masses.  For example, in Fig.~\ref{fig:NmassNLO} we display the graphs contributing to the nucleon mass at $\mc{O}(m_q)$, Fig.~\ref{fig:NmassNLO}(a), $\mc{O}(a)$, Fig.~\ref{fig:NmassNLO}(b) and $\mc{O}(m_q^{3/2})$, Fig.~\ref{fig:NmassNLO}(c) and Fig.~\ref{fig:NmassNLO}(d).
%
%
The expressions for the quark mass dependence of the nucleon and delta masses can be found for example in Ref.~\cite{Tiburzi:2005na}, and the lattice spacing dependent corrections to these masses for Wilson fermions in the isotropic limit can be found in Refs.~\cite{Beane:2003xv,Tiburzi:2005vy}.

At $\mc{O}(am_q)$, there are two types of mass corrections, those from loop graphs from the lattice spacing dependent operators in Eq.~\eqref{eq:NTphiLag} and tree level terms from operators in the $\mc{O}(am_q)$ Lagrangian.  These corrections will be identical in form to those present in the isotropic limit, but with different numerical values due to the new anisotropic operators which do not explicitly break the hypercubic symmetry.  The loop graphs for the nucleon and delta mass corrections are depicted in Ref.~\cite{Tiburzi:2005vy} in Figs.~1 and 2 respectively.  In the isospin limit, there are two tree level operators for the nucleon and deltas each,
\begin{align}
	\mc{L}_{NT\phi}^{(am)} =&
		-n_{MW} \bar{N} N \tr(\mc{M}_+) \tr(\mc{W}_+)
		-n_{MW}^\xi \bar{N} N \tr(\mc{M}_+) \tr(\mc{W}^\xi_+)
	\nonumber\\&
		+t_{MW} \bar{T}_\mu T_\mu \tr(\mc{M}_+) \tr(\mc{W}_+)
		+t_{MW}^\xi \bar{T}_\mu T_\mu \tr(\mc{M}_+) \tr(\mc{W}^\xi_+)\, ,
\end{align}
which also function as counterterms for divergences from the above mentioned loop graphs.  As with the mesons, we must consider the $\mc{O}(a^2)$ effects to see the first explicit breaking of the anisotropy.  Even in the isotropic limit, the lattice spacing corrections to the baryons at this order proved to be quite interesting, with the presence of the first operators which arise from the $\mc{O}(4)$ breaking down to the hypercubic group at this order~\cite{Tiburzi:2005vy}.  There are several generic $\mc{O}(a^2)$ operators which do not explicitly break the hypercubic group,
\begin{align}
	\mc{L}_{NT\phi}^{a^2} =& 
		-n_{WW} \bar{N} N \tr(\mc{W}_+) \tr(\mc{W}_+)
		-n_{WW}^\xi \bar{N} N \tr(\mc{W}^\xi_+) \tr(\mc{W}^\xi_+)
	\nonumber\\&
		+t_{WW} \bar{T}_\mu T_\mu \tr(\mc{W}_+) \tr(\mc{W}_+)
		+t_{WW}^\xi \bar{T}_\mu T_\mu \tr(\mc{W}^\xi_+) \tr(\mc{W}^\xi_+)
	\nonumber\\&
		-\bar{n}_{WW}^\xi \bar{N} N \tr(\mc{W}_+) \tr(\mc{W}^\xi_+)
		+\bar{t}_{WW}^\xi \bar{T}_\mu T_\mu \tr(\mc{W}_+) \tr(\mc{W}^\xi_+)\, ,
\end{align}
where the last two operators vanish for the $\mc{O}(a)$ improved action and the first two survive the isotropic limit.  The $\mc{O}(4)$ breaking operators appear at this order in the heavy baryon Lagrangian because in the heavy baryon theory, there is an additional vector, the four-velocity of the baryon one can use to construct operators.  For the mesons, the only vector is $\partial_\mu$ and so the $\mc{O}(4)$ breaking operators in the meson Lagrangian don't appear until $\mc{O}(a^2 p^4)$~\cite{Bar:2003mh}.  These heavy baryon $\mc{O}(4)$ breaking operators are~\cite{Tiburzi:2005vy}
\begin{equation}\label{eq:baryon_O4}
	\mc{L}_{NT\phi}^{\mc{O}(4)} = a_s^2 n_{4}\, \bar{N} v_\mu v_\mu v_\mu v_\mu N
		+a_s^2 t_{4}\, \bar{T}_\rho v_\mu v_\mu v_\mu v_\mu T_\rho \, .
\end{equation}
Similar to these, the $\mc{O}(3)$ breaking operators appear at this order,
\begin{align}\label{eq:baryon_O3}
	\mc{L}_{NT\phi}^{\mc{O}(3)} =\ 
		a_s^2 n_3\, \bar{N} 
		\bar{\d}^\xi_{\mu\nu}
		\bar{\d}^\xi_{\mu\nu}
		\bar{\d}^\xi_{\mu\nu}
		\bar{\d}^\xi_{\mu\nu}\ 
	 N 
		+a_s^2 t_3\, \bar{T}_\rho 
		\bar{\d}^\xi_{\mu\nu}
		\bar{\d}^\xi_{\mu\nu}
		\bar{\d}^\xi_{\mu\nu}
		\bar{\d}^\xi_{\mu\nu}\ 
		 T_\rho\, .
\end{align}
However, even though these two sets of operators explicitly break the $\mc{O}(4)$ and $\mc{O}(3)$ symmetries respectively, they do not lead to modifications of the dispersion relations until higher orders~\cite{Tiburzi:2005vy}, and thus function as mass corrections.  There is a subtlety of the heavy baryon Lagrangian related to reparameterization invariance (RPI)~\cite{Luke:1992cs}, which  constrains coefficients of certain operators in the heavy baryon Lagrangian.  For example, in the continuum limit, the LO kinetic operator of Eq.~\eqref{eq:NTphiLag} is related to a higher dimensional operator in such a way as to provide the correct dispersion relation,%
\footnote{In this equation, $D_\perp^2 = D^2 - (v\cdot D)^2$.}
\begin{equation}
 \mc{L} = \bar{N} v\cdot D N \longrightarrow
	\bar{N} v\cdot D N + \bar{N} \frac{D_\perp^2}{2M_N} N\, ,
\end{equation}
such that the energy of the non-relativistic nucleon is given by
\begin{equation}
	E = M_N + \frac{\mathbf{p}^2}{2M_N} + \dots
\end{equation}
For the anisotropic theory, this dispersion relation is then modified by an additional operator.  While in the continuum limit, RPI constrains the coefficient in front of the $1/M_N$ operator, the anisotropic action gives rise to modifications of this relation.
\begin{multline}
 \mc{L} = \bar{N} v\cdot D N + \bar{N} \frac{D_\perp^2}{2M_N} N \longrightarrow\
 \\
 \mc{L}^\xi = \bar{N} v\cdot D N 
	+\bar{N} \frac{D_\perp^2}{2M_N} N
	+a_t \sigma_E \bar{N} v\cdot u^\xi u^\xi \cdot D N
	+a_s \sigma_{KE} \bar{N} (u^\xi \cdot D)^2 N\, .
\end{multline}
The resulting nucleon dispersion relation from this action yields
\begin{multline}
E_N = M_N  + \frac{\mathbf{p}^2}{2M_N}  \longrightarrow\
	\\
	E_N (1+a_t \s_E)  = M_N(1+a_t \s_E) 
		-2N_f a_s ( \sigma_W +\s_W^\xi)
		+ \frac{\mathbf{p}^2}{2M_N} 
		+a_s \sigma_{KE}\, (E_N-M_N)^2 \, .
\end{multline}
For the $\mc{O}(a)$-improved action, the form of this dispersion relation stays the same with $\{a_t,a_s\} \rightarrow \{a_t^2, a_s^2\}$.  Comparing this dispersion relation to that of the pion, Eq.~\eqref{eq:pion_dispersion_Oa}, it is clear that even if the pion dispersion relation were tuned to be continuum like, the nucleon dispersion relation would still contain lattice spacing artifacts.  This is a simple consequence of the fact that the LECs in the heavy baryon Lagrangian (both the physical and unphysical) have no relation to those in the meson chiral Lagrangian.

%
%
\section{Discussion \label{sec:concl}}

We have developed the effective theory describing pions, nucleons and deltas at low energy, including the finite lattice spacings effects from an anisotropic Wilson lattice action. In particular, we focussed on the correction to the dispersion relations caused by the breaking of hypercubic symmetry.  The theory considered here is suited for the anisotropic Wilson lattices currently being generated.  Extensions to a mixed action case~\cite{Renner:2004ck,Bar:2002nr}, for example with domain-wall or overlap valence fermions and the Wilson sea fermions, both anisotropic, can be made with ease as in Refs.~\cite{Chen:2005ab,Chen:2006wf,Orginos:2007tw,Chen:2007ug}.  We have also highlighted some subtlety involving the Aoki-phase for anisotropic actions, which may be of importance in tuning the bare fermion masses in domain-wall and overlap fermions; for a fixed bare fermion mass and spatial lattice spacing, if the isotropic action is in the QCD-phase, the anisotropic theory may still be in the Aoki-phase.  This work will aid in the continuum extrapolation of correlation functions computed with anisotropic lattices.

%
%
\begin{acknowledgments}
We would like to thank Robert Edwards, Balint Joo, Kostas Orginos and David Richards for useful discussions and correspondence.  This research was supported in part by the U.S. Dept. of Energy under grant no. DE-FG02-93Er-40762.

\end{acknowledgments}

%
%

\end{document}